\documentstyle [11pt]{article}

\def\ber#1#2{\begin{equation}\begin{array}{#1}\displaystyle{#2}}
\def\ber#1{\begin{equation}\begin{array}{#1}\displaystyle}
\def\bernn#1#2{$$\begin{array}{#1}\displaystyle{#2}}

\def\eer#1{\end{array}\label{#1}\end{equation}}
\def\eernn{\end{array}$$}
\def\r#1#2{\noindent\hbox{\hbox to 24 pt{\hfil[#1]~}%
\vtop{\hsize = 12.5 truecm\noindent#2}}\vskip 5 pt\vfil}
\def\chap#1#2#3{\noindent\hbox{\hbox to 1.5 truecm{\hfil#1}%
\hbox to 14 truecm{~#2\leaders\hbox to 0.5 em{\hfil.\hfil}\hfill#3}}\par}
\def\cchap#1#2#3#4{\noindent\hbox{\hbox to 1.5 truecm{\hfil#1}%
\hbox to 14 truecm{~#2\hfil}}\par
\noindent\hbox{\hskip 1.5 truecm%
\hbox to 14 truecm{~#3\leaders\hbox to 0.5 em{\hfil.\hfil}\hfill#4}}\par}
\def\bbt{\bibitem}
\def\be{\begin{equation}}
\def\en{\end{equation}}
\def\ber{\begin{eqnarray}}
\def\enr{\end{eqnarray}}
\def\nmb{ \nonumber\\}
\def\d{\partial}
\def\rbr{\rbrack}
\def\lbr{\lbrack}
\def\rbrc{\rbrace}
\def\lbrc{\lbrace}
\def\ov{\over }
\def\tld{\tilde}

\def\MTR{Manin triple }
\def\MTRs{Manin triples }
\def\DLG{double Lie group }
\def\Tta{\Theta}
\def\sgm{\sigma}

\def\im{\imath}
\def\rh{\rho}
\def\lm{\lambda}
\def\Lm{\Lambda}

\begin{document}
\rightline{Landau Tmp/11/96.}
\rightline{December 1996}
\vskip 2 true cm
\centerline{\bf POISSON-LIE T-DUALITY IN N=2
SUPERCONFORMAL FIELD THEORIES.}
\vskip 2.5 true cm
\centerline{\bf S. E. Parkhomenko}
\centerline{Landau Institute for Theoretical Physics}
\centerline{142432 Chernogolovka,Russia}
\vskip 1 true cm
\centerline{\bf Abstract}
\vskip 0.5 true cm
 The supersymmetric generalization of Poisson-Lie T-duality
in superconformal WZNW models is considered.
It is shown that the classical N=2 superconformal WZNW models
posses a natural Poisson-Lie symmetry which allows to construct
dual $\sigma$- models.
\smallskip
\vskip 10pt
\centerline{\bf Introduction.}
Target space duality in string theory has attracted a considerable
attention in recent years because it sheds some light on
the geometry and symmetries of string theory. The well known
example of T-duality is  mirror symmetry in the Calaby-Yao
manifolds compactifications of the superstring. Duality symmetry
was first described in the context of toroidal compactifications
~\cite{BGS}. For the simplest case of single compactified dimension
of radius $R$, the entire physics of interacting theory is left
unchanged under the replacement $R\to \alpha /R$ provided one also
transforms the dilaton field $\phi \to \phi -\ln{(R/\sqrt{\alpha})}$
~\cite{AlO}. This simple case can be generalized to arbitrary toroidal
compactifications ~\cite{NarSW}. The T-duality symmetry
was later extended to the case of nonflat conformal backgrounds
possesing some abelian isometry (abelian T-duality) in ~\cite{Bu}.

 Of more recent history is the notion of non-abelian duality
~\cite{OsQ,GR}. The basic idea of ~\cite{OsQ} is to consider
a conformal field theory with non-abelian symmetry group.
The non-abelian duality did miss a lot of features characteristic
to the abelian duality. For example the non-abelian T-duality
transformation of the isometric $\sigma$- model on a group
manifold $G$ gives non-isometric $\sigma$- model on its Lie algebra
~\cite{OsQ,FrJ,FrT,CurZ}. As a result, it was not known how to perform
the inverse duality transformation to get back to the original model.
Indeed, while the original model on $G$ was isometric, which was
believed to be an essential condition for performing a duality
transformation, the dual one did not possess the $G$- isometry.

 A solution of this problem was proposed recently in ~\cite{KlimS}
where it was argued that the two theories are dual to each other
from the point of view of the so called Poisson-Lie T-duality.
In ~\cite{KlimS} a large class of new dual pairs of $\sigma$-models
associated with each Drinfeld double ~\cite{Drinf1} was constructed.
The main idea of the approach is to replace
the requirement of isometry by a weaker condition
which is the Poisson-Lie symmetry of the theory. This generalized
duality is associated with two groups forming a Drinfeld double
and the duality transformation exchanges their roles.

 The discussion in ~\cite{KlimS} was quite general.
In order to apply Pisson-Lie T-duality in superstring
theory one would like to know if there
are dual pairs of conformal and superconformal
$\sigma$-models. In particular, it would be intresting to
construct mutualy dual pairs of N=2 superconformal field theories.

The simple example of dual pair of conformal field theories
associated with the $O(2,2)$ Drinfeld double was presented
in work ~\cite{AlKT}. The supersymmetric generalization
of Poisson-Lie T-duality was considered in~\cite{Sfet}.
The present note is devoted to the construction of dual
pairs of N=2 superconformal WZNW models. In particular,
after brief review of the Manin triple construction of
N=2 superconformal WZNW models in the section 1,
we will show in the
section 2, that the classical N=2 superconformal WZNW models
posses very natural Poisson-Lie symmetry which we will use
to construct Poisson-Lie T-dual $\sigma$-models. In section 3 we will
apply the results of section 2 to N=2 superconformal
WZNW model associated with the Manin triple
$(sl(2,R)\oplus R, b_{+}, b_{-})$,
where $b_{\pm}$ are the Borel subalgebras of $sl(2,R)$
and construct its Poisson-Lie dual $\sigma$-model.

\vskip 10pt
\centerline{\bf1. The classical N=1 superconformal WZNW model.}

 In this section we briefly review a supersymmetric
WZNW (SWZNW) models using superfield formalism ~\cite{swzw}
and formulate
conditions that a Lie group should satisfy in order for its
SWZNW model to possess extended supersymmetry.

 We parametrize superworld sheet introducing the light cone
coordinates
$x_{\pm}$, and grassman coordinates $\Tta_{\pm}$.
The generators of supersymmetry and covariant
derivatives are
\be
Q_{\mp}= {\d \ov \d\Tta_{\pm}}+\im \Tta_{\pm}\d_{\mp},\
D_{\mp}= {\d \ov \d\Tta_{\pm}}-\im \Tta_{\pm}\d_{\mp}.
\label{1}
\en
They satisfy the relations
\be
\lbrc D_{\pm},D_{\pm}\rbrc= -\lbrc Q_{\pm},Q_{\pm}\rbrc= -\im 2\d_{\pm},\
\lbrc D_{\pm},D_{\mp}\rbrc= \lbrc Q_{\pm},Q_{\mp}\rbrc=
\lbrc Q,D\rbrc= 0,
\label{2}
\en
where the brackets $\lbrc,\rbrc$ denote the anticommutator.
The superfield of N=1 supersymmetric WZNW model
\be
G= g+ \im \Tta_{-}\psi_{+}+ \im \Tta_{+}\psi_{-}+
   \im \Tta_{-}\Tta_{+}F  \label{3}
\en
takes values in a Lie group ${\bf G}$.
We will assume that its Lie algebra ${\bf g}$
is endowed with ad-invariant nondegenerate inner
product $<,>$.

The inverse group element $G^{-1}$ is defined from
\be
 G^{-1}G=1 \label{4}
\en
and has the decomposition
\be
 G^{-1}= g^{-1}- \im \Tta_{-}g^{-1}\psi_{+}g^{-1}-
         \im \Tta_{+}g^{-1}\psi_{-}g^{-1}-
         \im \Tta_{-}\Tta_{+}g^{-1}(F+\psi_{-}g^{-1}\psi_{+}-
         \psi_{+}g^{-1}\psi_{-})g^{-1} \label{5}
\en
For physical reasons one has to demand the group ${\bf G}$
is a real manifold. Therefore it is convenient to consider ${\bf G}$
as a subgroup in the group of real or unitary matrixes i.e. one
has to impose  the following conditons on the matrix elements of
the superfield $G$:
\be
\bar{g}^{mn}=g^{mn},\
\bar{\psi}^{mn}_{\pm}= \psi^{mn}_{\pm},\
\bar{F}^{mn}= F^{mn} \label{6.r}
\en
or
\be
\bar{g}^{mn}=(g^{-1})^{nm},\
\bar{\psi}^{mn}_{\pm}= (\psi^{-1})^{nm}_{\pm},\
\bar{F}^{mn}= (F^{-1})^{nm}, \label{6.u}
\en
where we have used the following notations
\be
\psi^{-1}_{\pm}= -g^{-1}\psi_{\pm}g^{-1},\
F^{-1}= -g^{-1}(F+\psi_{-}g^{-1}\psi_{+}-
         \psi_{+}g^{-1}\psi_{-})g^{-1}. \label{6.not}
\en
In the following we will assume that the superfield $G$
satisfy (\ref{6.r}) i.e. the Lie group ${\bf G}$ is
a subgroup of the group of nondegenerate real matrixes.

The action of N=1 SWZNW model is given by
\ber
S_{swz}= \int d^{2}x d^{2} \Tta(<R_{+},R_{-}>)   \nmb
         -\int d^{2}x d^{2}\Tta dt
          <G^{-1}\frac{\d G}{\d t},\lbrc R_{-},R_{+}\rbrc>, \label{7}
\enr
where
\be
 R_{\pm}= G^{-1}D_{\pm}G.  \label{8}
\en
The classical equations of motion can be obtained by making a variation
of (\ref{7}):
\be
\delta S_{swz}= \int d^{2}x d^{2} \Tta
<G^{-1}\delta G,D_{-}R_{+}-D_{+}R_{-}-\lbrc R_{-},R_{+}\rbrc>
\label{9}
\en
Taking into account kinematic relation
\be
D_{+}R_{-}+D_{-}R_{+}= -\lbrc R_{+},R_{-}\rbrc \label{rel}
\en
we obtain
\be
 D_{-}R_{+}=0
\label{10}
\en

The action (\ref{7}) is invariant under the super-Kac-Moody
and N=1 superconformal
transformations ~\cite{swzw}.

 In the following we will use
supersymmetric version of Polyakov-Wiegman formula ~\cite{PW}
\be
S_{swz}[GH]= S_{swz}[G]+ S_{swz}[H]+ \int d^{2}x d^{2}\Tta
             <G^{-1}D_{+}G,D_{-}HH^{-1}>.  \label{11}
\en
It can be prooved as in the non supersymmetric case.

 In works ~\cite{QFR2,QFR3,QFR} supersymmetric WZNW models which admit
extended supersymmetry were studied and correspondence between extended
supersymmetric WZNW models and finite-dimensional Manin triples was
established in ~\cite{QFR3,QFR}.
By the definition ~\cite{Drinf1},
a \MTR $({\bf g},{\bf g_{+}},{\bf g_{-}})$
consists
of a Lie algebra ${\bf g}$, with nondegenerate invariant inner product
$<,>$ and isotropic Lie subalgebras ${\bf g_{\pm}}$ such that
${\bf g}={\bf g_{+}}\oplus {\bf g_{-}}$ as a vector space.
The corresponding Sugawara construction of N=2 Virasoro superalgebra
generators was given in ~\cite{QFR3,QFR,GETZ}.

 To make a connection between Manin triple construction of
~\cite{QFR3,QFR} and approach of ~\cite{QFR2} based on complex
structures on Lie algebras the following comment is relevant.

 Let ${\bf g}$ be a real Lie algebra and $J$ be a complex structure
on the vector space ${\bf g}$. $J$ is referred to as a complex structure
on a Lie algebra ${\bf g}$ if $J$ satisfies the equation
\be
\lbr Jx,Jy \rbr-J\lbr Jx,y \rbr-J\lbr x,Jy \rbr=\lbr x,y \rbr \label{12}
\en
for any elements $x, y$ from ${\bf g}$.
Suppose the existence of a nondegenerate invariant inner product $<,>$ on
${\bf g}$ so that the complex structure $J$ is skew-symmetric with
respect to $<,>$. In this case it is not difficult to establish the
correspondence between complex \MTRs and complex structures on
Lie algebras. Namely, for each complex \MTR
$({\bf g},{\bf g_{+}},{\bf g_{-}})$
exists a canonic complex structure on the Lie algebra ${\bf g}$ such
that subalgebras ${\bf g_{\pm}}$ are its $\pm \im$ ei\-gen\-spa\-ces.
On the other hand, for each real Lie algebra ${\bf g}$
with nondegenerate invariant inner product and
skew-symmetric complex structure $J$ on this algebra one can
consider the complexification ${\bf g^{C}}$ of ${\bf g}$. Let
${\bf g_{\pm}}$ be $\pm \imath$ eigenspaces of $J$
in algebra ${\bf g^{C}}$ then $({\bf g},{\bf g_{+}},{\bf g_{-}})$
is a complex \MTR.
Moreover it can be prooved ~\cite{QFR3} that there exists one-to-one
correspondence between complex Manin triple endowed with a hermitian
conjugation (involutive antioutomorphysm) $\tau: {\bf g_{\pm}}\to
{\bf g_{\mp}}$ and the real Lie
algebra endowed with $ad$-invariant nondegenerate inner product $<,>$
and the complex structure $J$ which is skew-symmetric with respect
to $<,>$. Therefore we can use this conjugation to extract a compact
form from a complex Manin triple.

 If a complex structure on a Lie algebra is fixed then it defines the
second supersymmetry transformation ~\cite{QFR2}.

In this paper we concentrate on  N=2 SWZNW models based on
real Manin triples. The case of N=2 SWZNW models on
compact groups will be considered in near future.

\vskip 10pt
\centerline{\bf2. Poisson-Lie T-duality in N=2 superconformal WZNW model.}

 In this section we will describe the construction of Poisson-Lie T-dual
$\sgm$-models to N=2 SWZNW models.

 For the description of the Poisson-Lie T-duality in N=2 SWZNW we need
a Lie group version of Manin triple ~\cite{SemTian,LuW,AlMal}.
Let's fix some \MTR $({\bf g},{\bf g_{+}},{\bf g_{-}})$
and consider
\DLG $({\bf G},{\bf G_{+}},{\bf G_{-}})$ ~\cite{LuW}, where the
exponential groups ${\bf G}$, ${\bf G_{\pm}}$ correspond to Lie algebras
${\bf g}$, ${\bf g_{\pm}}$. Each element $g\in {\bf G}$ admits a
decomposition
\be
g= g_{+}g^{-1}_{-}  \label{13}
\en
For SWZNW model on the group ${\bf G}$ we obtain from
(\ref{13}) the decomposition for the superfield (\ref{4})
\be
G(z_{+},z_{-})= G_{+}(z_{+},z_{-})G^{-1}_{-}(z_{+},z_{-}) \label{14}
\en
Due to (\ref{14}), (\ref{11}) and the definition of Manin triple we
can rewrite the action (\ref{7}) for this model in the following form
\be
S_{swz}=-\int d^{2}x d^{2}\Tta <\rh^{+}_{+}, \rh^{-}_{-}>,  \label{15}
\en
where the superfields
\be
\rh^{\pm}= G^{-1}_{\pm}DG_{\pm} \label{16}
\en
correspond to the right invariant 1-forms on the groups ${\bf G_{\pm}}$.

 To generalize (\ref{13}) we have to consider the set $W$ of classes
${\bf G_{+}}\backslash {\bf G}/ {\bf G_{-}}$ and pick up a representative $w$ for
each class $[w]\in W$ ~\cite{AlMal}:
\be
{\bf G}= \bigcup_{[w]\in W} {\bf G_{+}}w{\bf G_{-}}=
         \bigcup_{[w]\in W} {\bf G_{w}}  \label{17}
\en
This formula means that there is the natural action of complex group
${\bf G_{+}\times G_{-}}$ on ${\bf G}$,
and the set $W$ parametrizes ${\bf G_{+}\times G_{-}}$-orbits
${\bf G_{w}}$.

 The corresponding generalization for the action (\ref{15}) is given by
\be
S_{swz}=-\int d^{2}x d^{2}\Tta <\rh^{+}_{+}, w\rh^{-}_{-}w^{-1}>
\label{18}
\en

 Following ~\cite{KlimS} we consider a variation of the action
(\ref{18}) for SWZNW model on the group ${\bf G}$ under the right
action ${\bf G_{+}}\times {\bf G_{-}}$. Let us concentrate at first
on the class of identity from (\ref{17}).
\ber
\delta S_{swz}=-\int d^{2}x d^{2}\Tta (<D_{+}X^{+},\rh^{-}_{-}>-
<D_{-}X^{-},\rh^{+}_{+}>)+   \nmb
\int d^{2}x d^{2}\Tta (<X^{+},\lbrc \rh^{+}_{+}, \rh^{-}_{-}\rbrc>-
<X^{-},\lbrc \rh^{-}_{-}, \rh^{+}_{+}\rbrc>),   \label{19}
\enr
where $X^{\pm}= G^{-1}_{\pm}\delta G_{\pm}$
. From (\ref{19}) we
obtain the Noetherian currents $\rh^{\pm}_{\pm}$ which satisfy
on extremals
\ber
D_{+}\rh^{-}_{-}+\lbrc \rh^{-}_{-}, \rh^{+}_{+}\rbrc^{-}=0,  \nmb
D_{-}\rh^{+}_{+}+\lbrc \rh^{-}_{-}, \rh^{+}_{+}\rbrc^{+}=0,   \label{20}
\enr
where the brackets $\lbrc,\rbrc$ correspond to Lie brakets on ${\bf g}$.
These equations can be joint into zero curvature equation for $F_{+-}$-
component of super stress tensor $F_{MN}$
\be
F_{+-}= \lbrc D_{+}+\rh^{+}_{+}, D_{-}+\rh^{-}_{-}\rbrc=0 \label{21}
\en
Using standard arguments of super Lax construction ~\cite{EvHol} one
can show that from (\ref{21}) it follows that the connection is flat
\be
F_{MN}=0,\ M, N= (+, -, +, -).  \label{22}
\en

 The equations (\ref{21}) are the supersymmetric
generalization of Poisson-Lie symmetry conditions ~\cite{KlimS}.
Indeed, Noetherian currents $\rh^{+}_{+}$, $\rh^{-}_{-}$
are generators of
${\bf G_{+}}\times {\bf G_{-}}$- action on ${\bf G}$, while the
structure constants in (\ref{21}) correspond to Lie algebra ${\bf g}$
which is Drinfeld's dual to ${\bf g_{+}\oplus g_{-}}$ ~\cite{Drinf2}.
Due to (\ref{22}) we may associate to each extremal surface
($G_{+}(x_{+}, x_{-}, \Tta_{+}, \Tta_{-})$,
$G_{-}(x_{+}, x_{-}, \Tta_{+}, \Tta_{-})$)
a mapping
$G(x_{+}, x_{-}, \Tta_{+}, \Tta_{-})$ from the super world sheet into
the group ${\bf G}$ such that
\be
L^{+}_{-}+\rh^{+}_{-}= L^{-}_{-}+\rh^{-}_{-}= 0,   \label{23}
\en
\be
L^{+}_{-}= L^{-}_{+}= 0,     \label{24}
\en
where $L^{\pm}$ are ${\bf g_{\pm}}$-components of the current
$L=DGG^{-1}$ and $G\in {\bf G}$.

One can rewrite (\ref{23}), (\ref{24}) as the equations in Drinfeld's
double of ${\bf G}$. Let
\be
{\bf D}= {\bf G}\times {\bf G} \label{25}
\en
For each pair of superfields $\Lm =(G_{1},G_{2})\in {\bf D}$ there is
the decomposition
\be
\Lm= (G_{+},G_{-})(G,G)= H\tld{H}, \label{26}
\en
where $H=(G_{+},G_{-})$, $\tld{H}=(G,G)$.
The equations (\ref{23}), (\ref{24}) can be rewritten in the form
\be
\ll D_{\pm}\Lm \Lm ^{-1}, E^{\mp}\gg= 0, \label{27}
\en
where $\ll, \gg$ is the natural inner product on Lie algebra
of ${\bf D}$ and the subspaces $E^{\pm}$
are given by
\be
E^{+}= (0,{\bf g}),\ E^{-}= ({\bf g}, 0) \label{28}
\en
The equations (\ref{27}) with appropriate choise of
mutualy orthogonal subspaces $E^{\pm}$ are the
supersymmetric generalization of corresponding equations
from ~\cite{KlimS}. In the case when the subspaces are
choosen in general position (which corresponds to nondegeneracy
of the bilinear form determining the Lagrangian of $\sgm$-model)
one can use the construction ~\cite{KlimS} to obtain the action
for Poisson-Lie T-dual $\sgm$-model. But in our case the subspaces
(\ref{28}) are not in general position and the bilinear form
determinig the Lagrangian of N=2 SWZNW model is singular
as it is easy to see from (\ref{15}). It makes impossible to
apply strightforward costruction ~\cite{KlimS}. Instead we will use
the method developed in ~\cite{AlKT}.

 Let us suppose that instead of (\ref{26}) we use the
decomposition
\be
\Lm= \tld{F}F, \label{29}
\en
where $F=(U_{+},U_{-})$, $\tld{F}=(U,U)$.
Taking into account (\ref{28}) we rewrite (\ref{27}) in the
following form
\be
R^{+}_{+}+\lm^{+}_{+}= R^{-}_{-}+\lm^{-}_{-}= 0,   \label{30}
\en
\be
R^{+}_{-}= R^{-}_{+}= 0,     \label{31}
\en
where $R^{\pm}$ are ${\bf g_{\pm}}$-components of the current
$R=U^{-1}DU$, $U\in {\bf G}$,
$\lm^{\pm}_{\pm}=D_{\pm}U_{\pm}U^{-1}_{\pm},\ U_{\pm}\in {\bf G_{\pm}}$.
In dual picture we should have an action for dual $\sgm$-model
on the group ${\bf G}$ and the action of ${\bf G}$ on itself
such that the Noetherian currents satisfy
zero curvature equation for $F_{+-}$-component of super stress
tensor taking values in Lie algebra ${\bf g_{+}}\oplus {\bf g_{-}}$
which is Drinfeld's dual Lie algebra to ${\bf g}$. But in view of the
constraint (\ref{31}) the action of dual $\sgm$- model has to be
contained corresponding Lagrange multipliers. Using the arguments
of ~\cite{AlKT} we can write the action of dual $\sgm$- model in the
following form
\be
\tld{S}_{swz}=-\int d^{2}x d^{2}\Tta (<\lm^{+}_{+}, \lm^{-}_{-}>+
<R_{+},\lm_{-}>+<\lm_{+},R_{-}>)
\label{32}
\en
It is easy to see from (\ref{32}) that the currents $\lm^{+}_{-}$,
$\lm^{-}_{+}$ play the role of the Lagrange multipliers (with values in
${\bf g_{+}}\oplus {\bf g_{-}}$).
The corresponding equations of motion include appart from
(\ref{30}), (\ref{31}) zero curvature equation
\be
\tld{F}_{+-}= \lbrc D_{+}-\lm^{+}_{+}+\lm^{-}_{+},
                    D_{-}-\lm^{-}_{-}+\lm^{+}_{-}\rbrc =0,
\label{33}
\en
where the brackets $\lbrc,\rbrc$ correspond to Lie brakets on
${\bf g_{+}\oplus g_{-}}$.
Excluding from (\ref{32}) all $\lm$'s exept the Lagrange
multipliers we obtain
\be
\tld{S}_{swz}=-\int d^{2}x d^{2}\Tta (<R^{+}_{+}, R^{-}_{-}>+
<R^{-}_{+},\lm^{+}_{-}>+<\lm^{-}_{+},R^{+}_{-}>)
\label{34}
\en

 Now we turn to the remainder adjacent classes from (\ref{17}).
The generalization of (\ref{21}), (\ref{23}), (\ref{24}) is
strightforward. In each class $[w]$ the Noetherian currents
$\rh^{+}_{w+}$ and $\rh^{-}_{w-}$
take values in the subspaces
${\bf g^{w}_{+}}= w^{-1}{\bf g_{+}}w\cup {\bf g_{+}}$
and
${\bf g^{w}_{-}}= w{\bf g_{-}}w^{-1}\cup {\bf g_{-}}$ correspondingly
, while the constraints $L^{-}_{w+}$ and $L^{+}_{w-}$ take values
in the complements ${\bf g}\setminus {\bf g^{w}_{+}}$ and
${\bf g}\setminus {\bf g^{w}_{-}}$:
\be
F_{w+-}=\lbrc D_{+}+\rh^{+}_{w+}, D_{-}+ \rh^{-}_{w-}\rbrc=0
\label{35}
\en
\be
L^{\pm}_{w\pm}+\rh^{\pm}_{w\pm}=0,   \label{36}
\en
\be
L^{\pm}_{w\mp}=0,     \label{37}
\en
The arguments we have used to obtain (\ref{34}) can be applied (with
relevant modifications) to each class $[w]$. Taking into account
(\ref{35}), (\ref{36}), (\ref{37}) we obtain the generalization
of (\ref{34})
\be
\tld{S}_{swz}=-\int d^{2}x d^{2}\Tta (<R^{+}_{w+}, R^{-}_{w-}>+
<R^{-}_{w+},\lm^{+}_{w-}>+<\lm^{-}_{w+},R^{+}_{w-}>),
\label{38}
\en
where $R^{+}_{w+}$ and $R^{-}_{w-}$ take values in the same subspaces
like the currents $\rh^{+}_{w+}$ and $\rh^{-}_{w-}$,
$R^{-}_{w+}$ and $R^{+}_{w-}$ take values
in the complements ${\bf g}\setminus {\bf g^{w}_{+}}$ and
${\bf g}\setminus {\bf g^{w}_{-}}$ correspondingly.

\vskip 10pt
\centerline{\bf3. Poisson-Lie T-dual $\sigma$-model to
N=2 SWZNW model on $\bf SL(2,R)\times R$.}

 The Lie algebra of the group ${\bf G}=SL(2,R)\times R$ has the basis
\be
e_{0}=\left(\begin{array}{cc}
                 1&0\\
                 0&0
                 \end{array}\right),\
e_{1}=\left(\begin{array}{cc}
                 0&1\\
                 0&0
                 \end{array}\right)
\label{41}
\en
\be
e^{0}=\left(\begin{array}{cc}
                 0&0\\
                 0&-1
                 \end{array}\right),\
e_{1}=\left(\begin{array}{cc}
                 0&0\\
                 1&0
                 \end{array}\right)
\label{42}
\en
Note that both sets of generators (\ref{41}), (\ref{42}) span
the Borelian subalgebras ${\bf b_{-}}$, ${\bf b_{+}}$
correspondingly and they are maximaly isotropic
with respect to the non-degenerate
invariant inner product defined by the brackets
\be
<e_{i},e^{j}>= \delta_{i}^{j}. \label{43}
\en
Hence we have \MTR $({\bf g},{\bf b_{+}},{\bf b_{-}})$.
Let ${\bf B_{\pm}}=\exp({\bf b_{\pm}})$.
The decomposition (\ref{17}) is given by Bruhat decomposition
\be
{\bf G}= {\bf G_{1}}\bigcup {\bf G_{w}}, \label{44}
\en
where
\ber
{\bf G_{1}}={\bf B_{+}}{\bf B_{-}},  \nmb
{\bf G_{w}}={\bf B_{+}}\left(\begin{array}{cc}
                                  0&1\\
                                  -1&0
                                  \end{array}\right){\bf B_{-}}
\label{45}
\enr

In the class of identity we parametrize the element
$g\in {\bf G_{1}}$
by the matrix
\be
g=\left(\begin{array}{cc}
            1&0\\
            u&1
            \end{array}\right)
      \left(\begin{array}{cc}
                1&v^{-1}\\
                0&1
            \end{array}\right)
      \left(\begin{array}{cc}
                \exp(a_{+})&0\\
                0&\exp(a_{-})
            \end{array}\right).
\label{46}
\en
The classical constraint $R^{-}_{+}=0$
takes the form
\ber
D_{+}a_{+}-v^{-1}D_{+}u=0, \nmb
D_{+}(u+v)=0,
\label{47}
\enr
the classical constraint $R^{+}_{-}=0$
takes the form
\ber
D_{-}u=0, \nmb
D_{-}a_{-}+v^{-1}D_{-}u=0
\label{48}
\enr
Under these constraints, the remaining components of the currents
are
\ber
R^{+}_{+}=-2D_{+}\omega e^{0}-
\exp(2\phi)D_{+}ve^{1}, \nmb
R^{-}_{-}=2D_{-}\omega e_{0}-
\exp(-2\phi)v^{-2}D_{-}ve_{1},
\label{49}
\enr
where we have introduced new variables
\be
\phi=(a_{+}-a_{-})/2, \
\omega=(a_{+}+a_{-})/2
\label {50}
\en
so that the action in (\ref{34}) becomes
\be
\tld{S_{1}}=\int d^{2}x d^{2}\Tta (-4D_{+}\omega D_{-}\omega +
D_{+}\ln{v}D_{-}\ln{v})
\label{51}
\en
and thus describes two free scalar superfields $\omega$ and $\ln{v}$.

 In the class [w] we have an appart from constrants (\ref{47}),
(\ref{48}) additional constraints
\ber
\exp(a_{+}-a_{-})D_{+}u=0, \nmb
\exp(a_{-}-a_{+})v^{-2}D_{-}(u+v)=0. \label{52}
\enr
Under the constraints (\ref{47}), (\ref{48}), (\ref{52})
the action in (\ref{38}) becomes
\be
\tld{S}_{w}=-\int d^{2}x d^{2}\Tta 4D_{+}\omega D_{-}\omega
\label{53}
\en
and thus describes free scalar superfield $\omega$.


\vskip 10pt
\centerline{\bf ACKNOWLEDGEMENTS}
\frenchspacing
 I'm very gratefull to A. Yu. Alekseev, B. L. Feigin, A. Kadeishvili,
I. Polubin and Y. Pugai
for discussions.
This work was supported in part by grant RFFI-96-02-16507,
INTAS-95-IN-RU-690.

\vfill
\end{document}